\documentclass[aps,prx,twocolumn,nopacs,superscriptaddress,floatfix,longbibliography]{revtex4-1}

\usepackage{graphicx}
\usepackage{verbatim}
\usepackage{mathtools}
\usepackage{amsmath}
\usepackage{sidecap}
\usepackage{epstopdf}
\usepackage{braket}
\usepackage{soul}
\usepackage{natbib}
\usepackage{physics}
\usepackage{float}

\begin{document}

\title{Mapping out the emergence of topological features in the highly alloyed topological Kondo insulators Sm$_{1-x}M_x$B$_6$ ($M$=Eu, Ce)}

\author{Yishuai Xu}
\author{Erica C. Kotta}
\affiliation{Department of Physics, New York University, New York, New York 10003, USA}
\author{M. S. Song}
\author{B. Y. Kang}
\author{J. W. Lee}
\author{B. K. Cho}
\affiliation{School of Materials Science and Engineering, Gwangju Institute of Science and Technology (GIST), Gwangju 61005, Korea}
\author{Shouzheng Liu}
\affiliation{Department of Physics, New York University, New York, New York 10003, USA}
\author{Turgut Yilmaz}
\author{Elio Vescovo}
\affiliation{National Synchrotron Light Source II, Brookhaven National Lab, Upton, New York 11973, USA}
\author{Jonathan D. Denlinger}
\affiliation{Advanced Light Source, Lawrence Berkeley National Laboratory, Berkeley, CA 94720, USA}
\author{Lin Miao}
\affiliation{School of Physics, Southeast University, Nanjing, 211189, China}
\author{L. Andrew Wray}
\email{lawray@nyu.edu}
\thanks{Corresponding author}
\affiliation{Department of Physics, New York University, New York, New York 10003, USA}

\begin{abstract}

SmB$_6$ is a strongly correlated material that has been attributed as a topological insulator and a Kondo insulator. Recent studies have found the topological surface states and low temperature insulating character to be profoundly robust against magnetic and non-magnetic impurities. Here, we use angle resolved photoemission spectroscopy to chart the evolution of topologically-linked electronic structure features versus magnetic doping and temperature in Sm$_{1-x}M_x$B$_6$ ($M$=Eu, Ce). Topological coherence phenomena are observed out to unprecedented $\sim$30\% Eu and 50\% Ce concentrations that represent extreme nominal hole and electron doping, respectively. Theoretical analysis reveals that a recent re-designation of the topologically inverted band symmetries provides a natural route to reconciling the persistence of topological surface state emergence even as the insulating gap is lost through decoherence.

\end{abstract}

\date{\today}
\maketitle

\twocolumngrid

%1. SmB6 intro: has received broad attention as a Kondo insulator (***old literature), a mixed-valent system (***), and more recently as a strongly correlated topological material (***cite reviews).  

%2. introduce topological scenario (inversion with a large energy scale at the X-point. for the easy-cleave [001] surface, this inversion projects to the $\overline{\Gamma}$-point and $\overline{X}$-point, meaning that topological surface states must lie between these momentum coordinates and the non-inverted $\overline{M}$-point [see Fig. 1(a)].

%3. recent studies have shown insulating character (***) and topological surface states (***) are remarkably robust against impurities - even magnetic. ***cite Kondo hole literature from Lin's paper on why Kondo insulators are not robust

%4. Here we.... (2 sentences describing findings.  This is the end of the first paragraph.

% a paragraph describing SmB6
\begin{SCfigure*}
\centering
    \includegraphics[width=11cm]{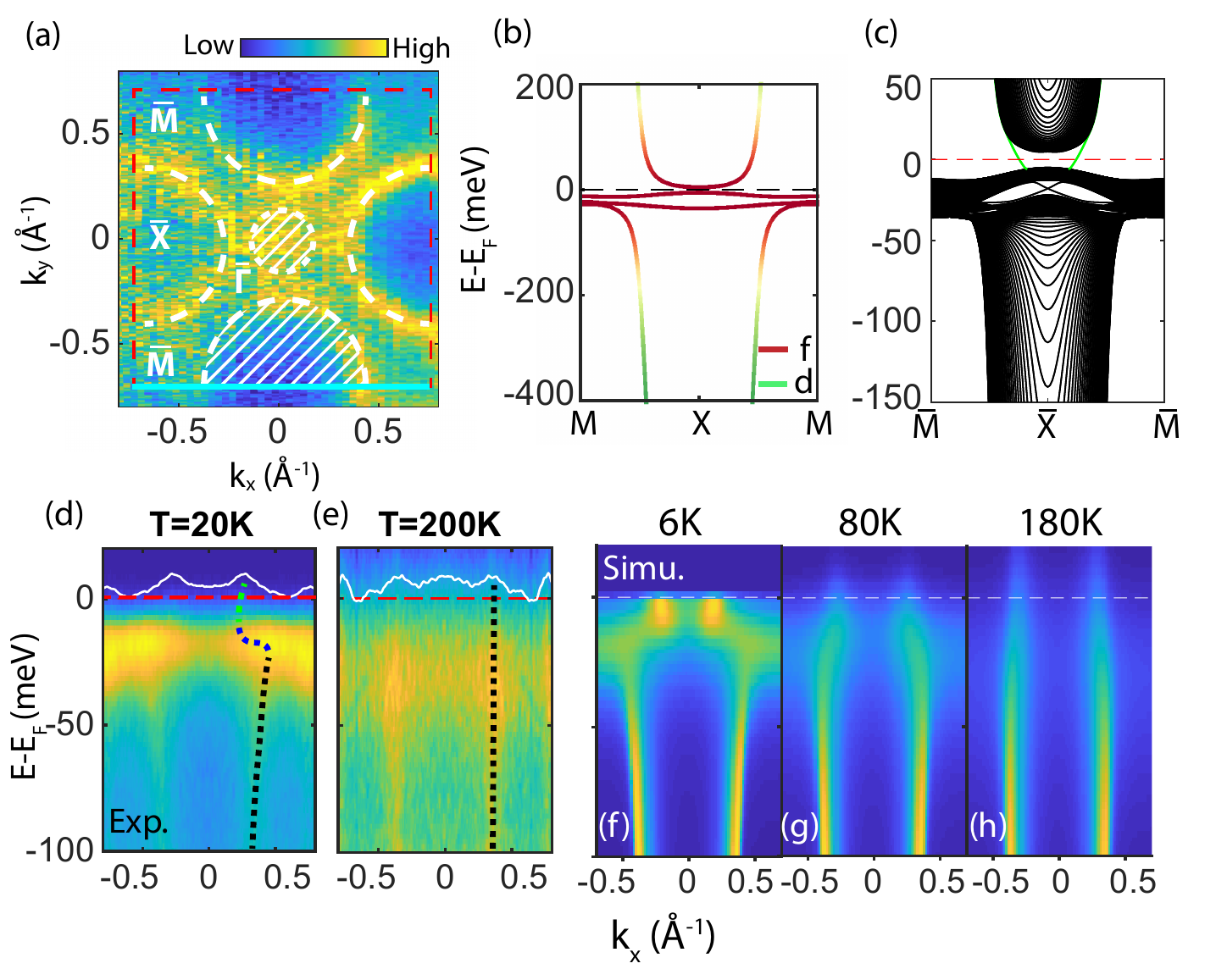} 
    \caption{Topological band structure of SmB$_6$. (a) A constant energy ARPES map of Ce$_{0.1}$Sm$_{0.9}$B$_{6}$ in the k$_z$$\sim$0 plane (h$\nu$=70 eV photon energy) is shown at binding energy $E_{B}=40$ meV, just beneath the 4$f$ bands.  Symmetry inversion regions surrounding the $\bar{X}$ and $\bar{\Gamma}$ point are delineated by the outer contour of the 5$d$ band, and have been highlighted in white. The $\bar{\Gamma}$ point 5$d$ Fermi surface is not visible due to the choice of photon energy, and the $\bar{\Gamma}$ point symmetry inversion region is shown with reduced size for visual clarity. (b) The bulk band structure from a tight binding model of SmB$_6$ is color coded to show orbital symmetry. (c) Simulated two-dimensional band structure of a slab configuration with $N=80$ unit cells stacked along the z-axis. In-gap topological surface states are indicted by the green solid line. (d-e) Experimental ARPES spectrum along the $\bar{M}-\bar{X}-\bar{M}$ cut of Ce$_{0.3}$Sm$_{0.7}$B$_{6}$ at T=20 and 200K. The intensity maximum as a function of binding energy is traced with the dashed line. (f-h) Simulated spectral functions at T=6, 80, and 200K, projected to the top 3 unit cells of the crystal.}
    \label{fig:fig1}
\end{SCfigure*}

The compound SmB$_6$ has been of great ongoing research interest as a Kondo insulator \cite{Menth_1969,Cohen_1970}, a mixed-valent system \cite{Kasuya_1994,ButchPressure,Skinner_2019} and as the prototypical example of a strongly correlated topological insulator \cite{Dzero2010,Dzero2012,XDai2013,Dzero_review_2016,AllenReview}. A topologically nontrivial interpretation of the band structure is supported by angle resolved photoemission spectroscopy (ARPES) investigations \cite{Neupane_2013,Jiang_2013,Denlinger_2013,denlinger2014temperature,Yoshiyuki_2019,AllenReview}, and is consistent with measurements of surface state spin texture \cite{Xu_2014_spinARPES}. However, the material remains enigmatic in many ways. Understanding of the low energy electronic band structure is fragmentary, and the symmetry character of the topological gap has recently been reattributed \cite{VojtaG8PRL, SigristG8PRL, VojtaG8PRB,SundermannG8,HeldSmB6DMFT}. Moreover, studies have shown that the bulk insulating character and surface states are phenomenally robust and can persist in alloys doped at the 10s of percent level, even with impurities that are nominally charge-donating and magnetic \cite{impurity1,impurity2,ImpurityPNAS,miao2020robust}. Together, these factors raise the fundamental question of how the electronic structure should be conceptualized at the boundary to the topological regime. Here we report ARPES measurements of band coherence phenomena in magnetically doped Sm$_{1-x}$M$_{x}$B$_6$ (M=Eu, Ce), tracking the band evolution and emergence of surface states as a function of both temperature and sample composition. Theoretical analysis reveals that the recent symmetry attribution of the topological gap enables a framework for understanding the large doping, temperature, and energy scales of coherence effects in the band structure, which greatly exceed expectations based on the size and temperature dependence of the insulating gap.

%The observed doping and temperature dependence are further discussed to contextualize the highly unusual resistance to impurities exhibited by the SmB$_6$ electronic structure.

% a paragraph about SmB6

%Paragraph 2:
%-start with panel b-c, describing what the f/d and surface state band structure look like along the $\overline{X}-\overline{M}$ axis of momentum space.

%-then go to panel a, because the reader will understand why this reveals the symmetry inversion regions

%-then talk about thermal loss of coherence - maybe even summarize the f-electron self energy at which coherence is lost (very briefly incorporate key details of model)

%- end with the expectation this gives us as coherence is lost in doped systems (zig-zag dispersion)

For the easy-cleave [001] surface, the topological band inversion of SmB$_6$ projects to the two dimensional (2D) $\overline{\Gamma}$-point and $\overline{X}$-point, meaning that topological surface states must lie between these momentum coordinates and the non-inverted $\overline{M}$-point [see Fig. 1(a)]. Band structure is relatively easy to observe along the $\overline{M}-\overline{X}-\overline{M}$ momentum axis \cite{miao2020robust, Denlinger_2013}, and can be understood from a tight binding model that includes one highly dispersive band derived from the Sm 5$d$ orbital, crossing the three j=5/2 4$f$ bands (Fig. \ref{fig:fig1}(b)). The positive parity eigenvalue of the 5$d$ band introduces a symmetry inversion at the 3D X-point, leading to a gapless surface state (green line) that appears inside the 5$d$/4$f$ hybridization gap (see Fig. \ref{fig:fig1}(c) slab calculation). Within the surface-sensitive ARPES spectral function, the surface state dispersion is only highly visible very close to the Fermi level (Fig. \ref{fig:fig1}(f)), and results in a zig-zag intensity contour that can be traced by applying a single-band fit to low temperature ARPES data (Fig. \ref{fig:fig1}(d)). Raising the temperature greatly increases the energy width of the 4$f$-bands \cite{denlinger2014temperature}, causing them to vanish from the apparent band dispersions (Fig. \ref{fig:fig1}(d-h)). The resulting spectrum shows only the 5$d$ band, which has a larger Fermi momentum than the low temperature surface states, and an enhanced group velocity due to the loss of 4$f$ hybridization.

%Paragraph 3:
%-then insert technique paragraph

Single crystals of Sm$_{1-x}$M$_x$B$_6$ (M= Eu, Ce) were prepared by the alumina flux method, and details of the growth method are described in Ref. \cite{miao2020robust}. High-resolution ARPES measurements were performed at beamline 4.0.3 at the Advanced Light Source and ESM beamline at NSLS-II, with a base pressure better than $5 \times 10^{-11}$ Torr. Both measurements were performed at $hv=70$ eV corresponding to a bulk $\Gamma$-plane (k$_z$=0) of the cubic Brillouin zone. X-ray diffraction characterization confirms single phase growth with lattice parameters that evolve monotoncially with doping [see supplemental material (SM) \cite{SM}]. Extensive sample characterization can also be found in Ref. \cite{miao2020robust}, including measurements of dopant homogeneity, crystallographic structure, impurity multiplet states, surface stability, and the 2D nature of surface states. All Eu alloyed samples retain a Kondo insulating resistivity trend \cite{Yeo_2012}, despite nominal hole doping by Eu as Eu$^{2+}$\cite{miao2020robust}. Ce alloys are nominally electron doped, and have relatively flat low temperature resistivity by 30$\%$ doping that resembles the Kondo metal end point of CeB$_6$. Detailed doping- and temperature-resolved measurements were performed with an in-house ARPES instrument equipped with a He lamp at photon energy $hv=21.2$ eV, for which the 2D $\overline{X}$-point is observed at $k_z\sim0.30 \AA^{-1}$ \cite{Denlinger_2013}. Momentum space alignment was performed in a fixed sample geometry, using a DA30 electron deflector analyzer. The base pressure was better than $1 \times 10^{-10}$ Torr for the He lamp ARPES system. All samples were cleaved in situ $T \lesssim 20K$, and all ARPES data were obtained within 12 hours after cleavage. The tight binding model is based on Ref. \cite{Baruselli_2014}, with details described in the SM \cite{SM}. We note that although the binding energy of the Dirac point can vary depending on surface modeling parameters, the ARPES spectral function is relatively insensitive to this factor \cite{SM}.

%Paragraph 4:
%-Note that the He lamp images look like $k_z$=0, but are actually believed to be $k_z$=***, near the edge of the 5d electron pocket. Talk about $k_z$ resolution and the underlying DOS distribution, and the fitting approach.
% Note for the kz ~0 curve, the x-axis has been re-scaled by a factor of 0.85

\begin{figure}
    \centering
    \includegraphics[width=8.7cm]{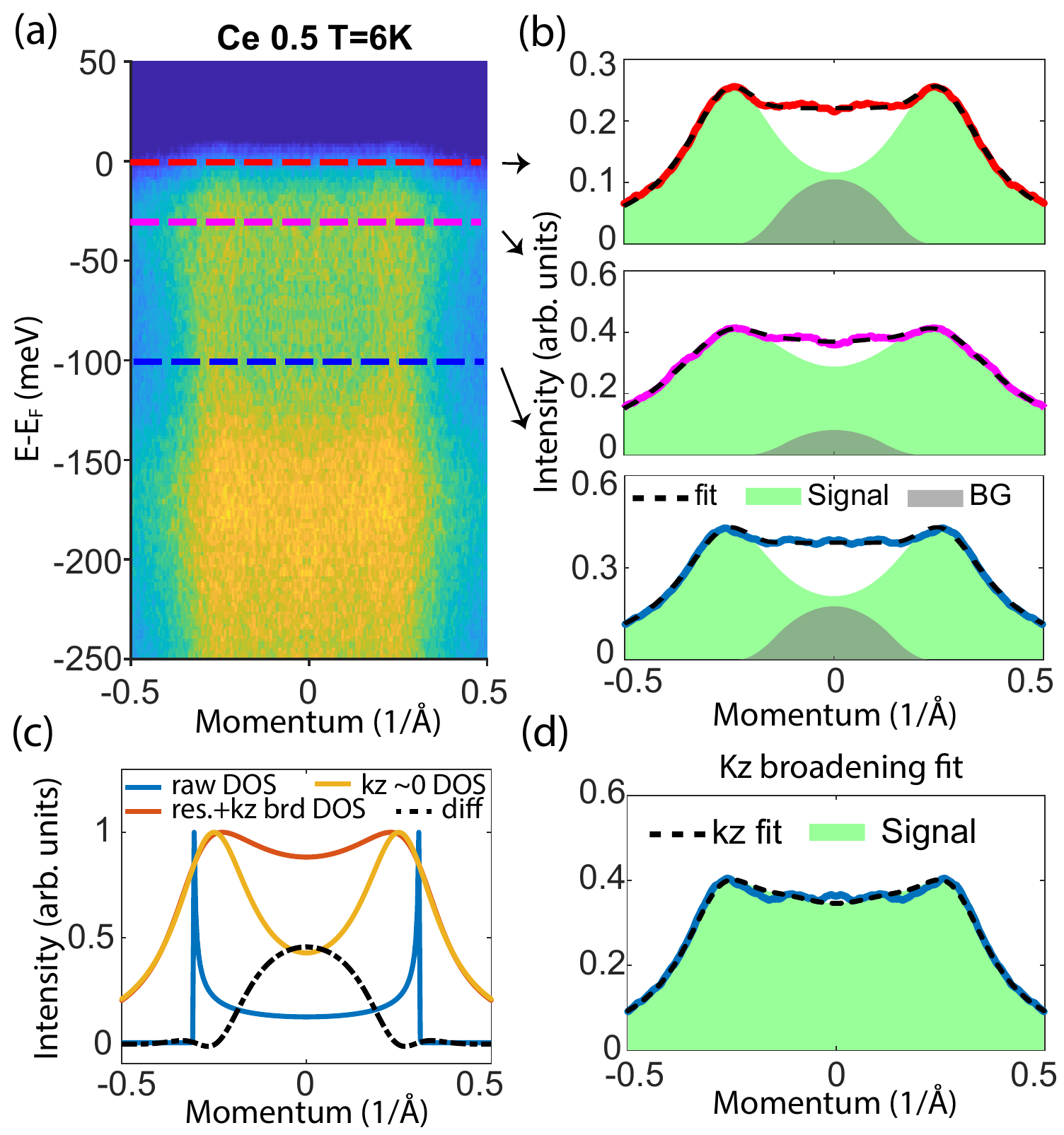}
    \caption{Fits of He lamp ARPES data. (a) The symmetrized ARPES spectrum along $\bar{M}-\bar{X}-\bar{M}$ obtained by He lamp for Ce$_{0.5}$Sm$_{0.5}$B$_{6}$. (b) Momentum distribution curves at energy $E-E_F$=-100, -30 and $0$ meV are overlaid with a two-component fit decomposition. The fit includes (green shaded area) two identical Lorentz peaks and (gray shaded area) a negative-curvature parabolic background, convoluted by experimental resolution. (c) A decomposition of the 5$d$ momentum distribution curve at $E-E_F$=-100 meV, showing (red) the fully modeled lineshape obtained from factoring in k$_z$ resolution, (blue) the underlying k$_z$-projected DOS, (orange) an 2-Lorentzian fit of the outer intensity contour, which resembles a k$_z$=0 spectrum, and (black dashed) the difference between the fully modeled lineshape and the 2-Lorentzian approximation. This difference resembles a negative-curvature parabola, and motivates the two-component fitting approach in panel (b).}
    \label{fig:fig2}
\end{figure}

An example of low temperature He lamp ARPES data from a highly alloyed SmB$_6$ sample (Sm$_{0.5}$Ce$_{0.5}$B$_6$, referenced as Ce 0.5) along the $\bar{M}-\bar{X}-\bar{M}$ direction is presented in Fig. \ref{fig:fig2}(a), and requires some contextual analysis to correctly interpret. The image shows the 5$d$ band dispersing through less visible non-dispersive 4$f$-derived features. The outer contour of the 5$d$ band is clearly visible at $k_x\sim\pm0.3 \AA^{-1}$, closely resembling the k$_z$=0 plane synchrotron data in Fig. \ref{fig:fig1}(e). This is surprisingly incompatible with the nominal surface-normal momentum of $k_z\sim0.30 \AA^{-1}$, where one expect to see the very edge of the 5$d$ electron pocket. Moreover, strong background intensity is present at smaller in-plane momenta ($k_x\sim0$). Both of these effects can be understood by considering the limited $k_z$ resolution of ARPES. In the absence of z-axis momentum resolution, the spectral function is expected to reflect the k$_z$-projected density of states (DOS) profile of the band (blue curve in Fig. \ref{fig:fig2}(c)), which has pronounced peaks at the outer band contour ($k_x\sim 0.3 \AA^{-1}$), corresponding to the band location in the k$_z$=0 plane. 

Modeling the $k_z$ resolution function as a Lorentzian (half-width of $\delta k_z$=0.57$\AA^{-1}$) causes these intense k$_z$$\sim$0 DOS peaks to show up as the dominant spectral feature, while photoemission from k$_z$ values near the nominal $k_z\sim0.30 \AA^{-1}$ plane appears only as background.  The resulting intensity profile closely matches experimental data (Fig. \ref{fig:fig2}(d)). Further contextual discussion of the ARPES $k_z$ resolution function can be found in the SM \cite{SM}. Note that the data are left-right symmetrized to restore the reflection symmetry of the spectral function, which is otherwise lost due to matrix elements associated with the geometry of the measurement. For continuum-like spectral features and highly disordered systems, matrix element effects can cause band dispersions to appear to differ slightly on the positive and negative momentum axes \cite{miao2018npj}. For fitting purposes, this $k_z$ resolution limited spectral function can be approximated by treating the outer edge of the spectral feature as a single $k_z\sim0.0$ band, and filling in the small-momentum background with a downward-opening parabola (see comparison in Fig. \ref{fig:fig2}(c)). Single band fits employing this simplified method are shown at multiple binding energies in Fig. \ref{fig:fig2}(b).

\begin{SCfigure*}
\centering
    \includegraphics[width=12cm]{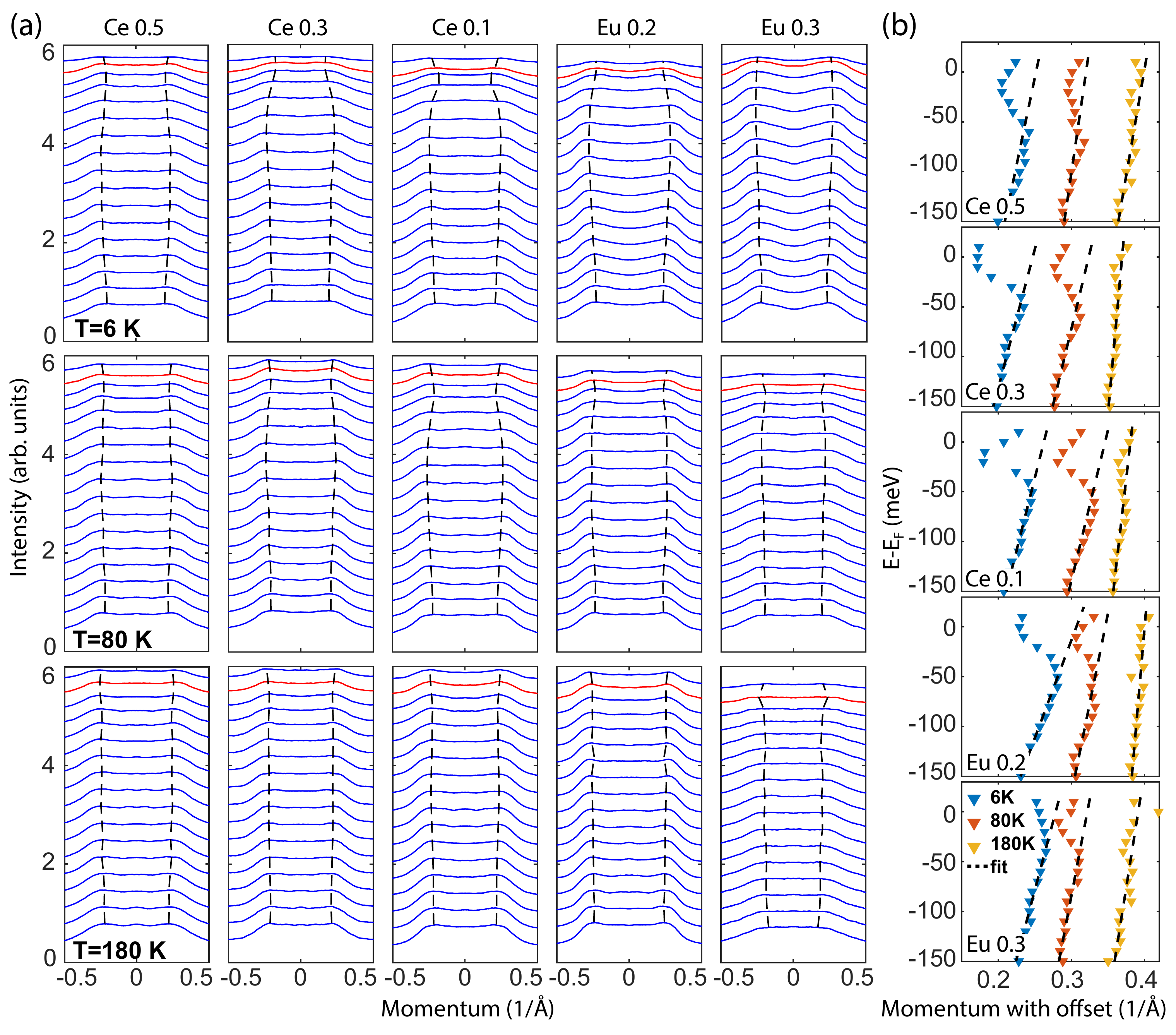} 
    \caption{Raw ARPES data and fitted dispersions. (a) Momentum distribution curves within 150 meV of the Fermi level are plotted with a 10 meV spacing for each doping series, at temperatures T=6, 80, and 180K. The Fermi level is indicated with a red curve, and a black dashed line traces the fitted peak positions. (b) Peak positions are plotted as a function of binding energy for each the doping series at T=6, 80 and 180K. Linear fits for the 5$d$ dispersion are obtained from the -150 to -50 meV energy window, and shown as black dashed lines. \newline \newline \newline \newline \newline \newline \newline \newline}
    \label{fig:fig3}
\end{SCfigure*}

%\begin{figure*}
%    \includegraphics[width=15cm]{Figure3.pdf}
%    \caption{Raw ARPES data and fitted dispersions. (a) Momentum distribution curves within 150 meV of the Fermi level are plotted with a 10 meV spacing for each doping series, at temperatures T=6, 80, and 180K. The Fermi level is indicated with a red curve, and a black dashed line traces the fitted peak positions. (b) Peak positions are plotted as a function of binding energy for each the doping series at T=6, 80 and 180K. Linear fits for the 5$d$ dispersion are obtained from the -150 to -50 meV energy window, and shown as black dashed lines.}
%    \label{fig:fig3}
%\end{figure*}

%Paragraph 5:
%-describe data in Fig. 3
%-after describing data, relate to Fig. 1 expectation

\begin{figure}
    \centering
    \includegraphics[width=7cm]{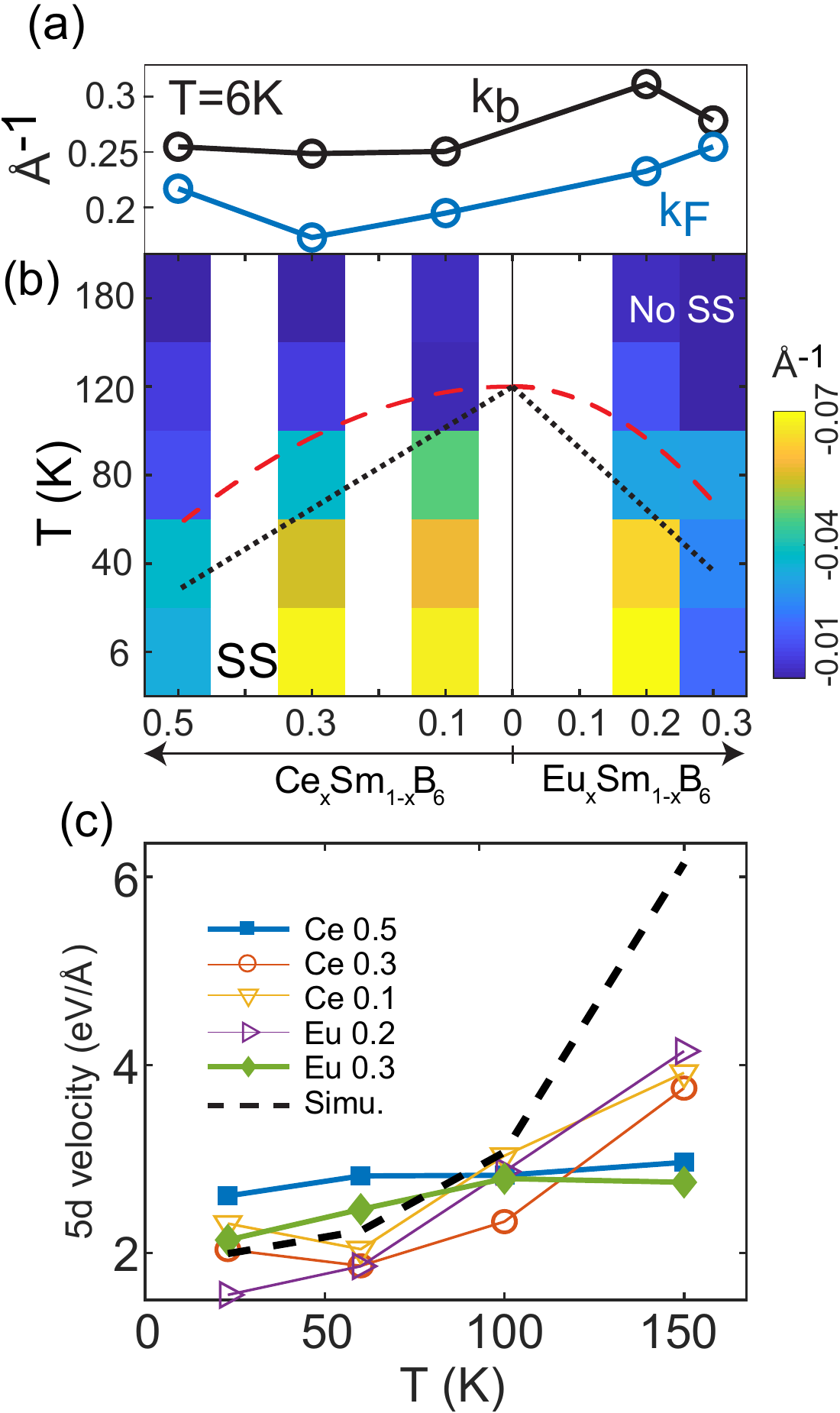}
    \caption{Identifying the topological Kondo cross-over. (a) The Fermi momentum observed at T=6K (blue) for each sample is compared with (black) the extrapolated T=6K Fermi momentum of the 5$d$ bulk band from Fig. \ref{fig:fig3}(b). (b) A map of surface state emergence, showing the maximum deviation between the fitted band contour and the extrapolated linear dispersion of the 5$d$ band near the Fermi level. Red and black dashed lines show expectations for surface state emergence based on treating the impurity scattering as a contribution to (black) imaginary and (red) real self energy, as discussed in the text. (c) The fitted 5$d$ band velocity in the -150 to -50 meV energy range, as a function of temperature. To reduce noise, 2-point averaging has been applied along the temperature axis. }
    \label{fig:fig4}
\end{figure}

\begin{figure}
    \centering
    \includegraphics[width=8.7cm]{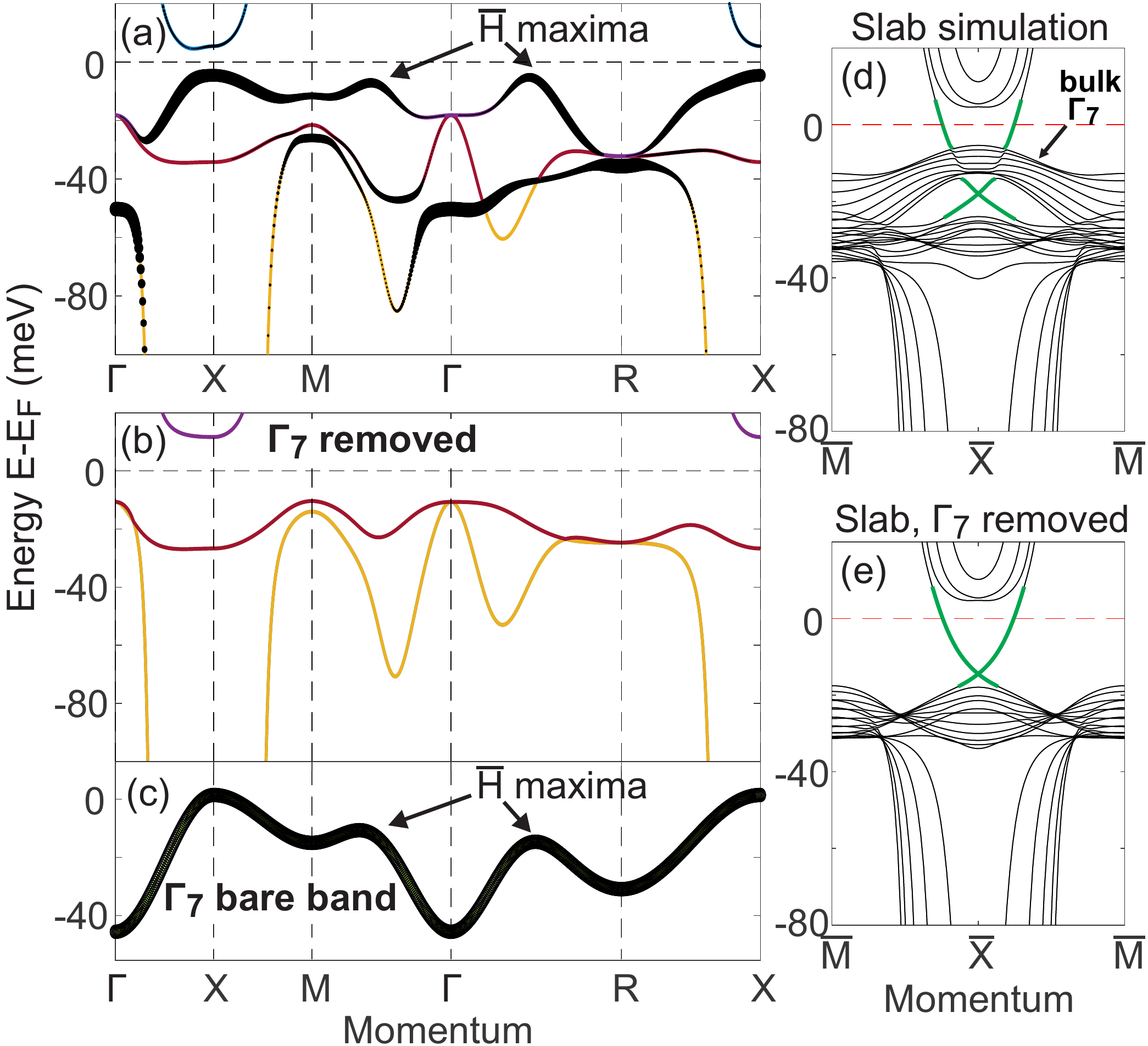}
    \caption{Distinct gaps for topology and conductivity. (a) Tight binding model band structure. Occupancy of $\Gamma_7$ orbital is shown by the thickness of a black overlay. Arrows indicate local maxima that roughly coincide with the 2D $\overline{H}$ point. The 2D $\overline{H}$ point is found at [0.25 0.25] in reciprocal lattice units, and is projected to from the midpoints of the $\Gamma$-M and $\Gamma$-R momentum segments. (b) The same model is shown with the $\Gamma_7$ states removed. (c) The $\Gamma_7$ dispersion is shown with all other orbitals removed. (d) The bulk band structure is shown from a 10-layer slab calculation. (e) The slab calculation is shown with the $\Gamma_7$ orbitals removed.}
    \label{fig:fig5}
\end{figure}

Measurements performed with the He lamp ARPES system as a function of temperature across the Eu and Ce doping series are shown in Fig. \ref{fig:fig3}. For all samples, the dispersions at high temperature (T=180K) show a mostly smooth linear dispersion, when fitted as described above [see fit summaries in Fig. \ref{fig:fig3}(b)]. As temperature is lowered, a clear deviation from linearity can be observed in all samples at T$\leq$80K, and the resulting intensity profile conforms to the zig-zag pattern that is expected upon the emergence of surface states. The linear 5$d$ band dispersion remains visible at energies $E-E_F<-50$ meV beneath the Fermi level, however the group velocity (slope) of the band is reduced. At the T=6K base temperature, the more highly doped samples (Sm$_{0.5}$Ce$_{0.5}$B$_6$, and Sm$_{0.7}$Eu$_{0.3}$B$_6$ referenced as Ce 0.5 and Eu 0.3) show smaller deviations from the extrapolated 5$d$ band dispersion.

%Paragraph 6:
%-move to Fig. 4
%-how does the low temperature amplitude of $\delta k$ vary with doping - what fraction of the max amplitude does it drop to, and is the change step function-like?
%-how about the thermal onset?
%-note that higher Eu doping is prohibitively hard to fit?

%The emergence of the surface states relies on the opening of hybridization gap which takes place only when the f band is coherent.

%%%%% 5/13/2021 Continue reading from here! %%%%%%%%%%

The deviation of band momentum from the linear 5$d$ trajectory is tracked as a metric for identifying the emergence of surface states in Fig. \ref{fig:fig4}(b). The maximum deviation of band momentum ($\Delta k$) is shown as a function of temperature and doping. At low temperature, the amplitude of $\Delta k$ is relatively flat across the doping series from Ce 0.3 to Eu 0.2 with an average value of $\Delta k=-0.07 \text{\AA}^{-1}$. Outside of this range, we observe a sudden change to $-0.03 \text{\AA}^{-1}$ and $-0.02 \text{\AA}^{-1}$ for Ce 0.5 and Eu 0.3 respectively. The thermal onset of $\Delta k(T)$ behaves as a roughly linear function within each doping series, with an origin between T=120 and 180K for most samples, in keeping with the onset of surface states documented in ARPES measurements on the undoped compound.

The merging of the low temperature Fermi momentum with the extrapolated 5$d$ band dispersion at high doping does not necessarily indicate the loss of the topological surface state. Separately tracking the extrapolated 5$d$ Fermi momentum and the low temperature Fermi momentum in Fig. \ref{fig:fig4}(a) reveals that the near-intersection of these momenta is associated with different kinds of anomalies at the two ends of the doping series. In the case of Ce 0.5, it the Fermi momentum appears to converge towards the 5$d$ band, consistent with the loss of the topological band coherence, and with nominal electron doping from Ce promoting a larger 5$d$ electron pocket. However, in the case of Eu doping, the 5$d$ Fermi momentum of the Eu 0.3 sample is sharply reduced, suggesting that the surface state may still be very robust but is simply not greatly offset from the $k_z=0$ 5$d$ band trajectory.

An interplay is also observed between the appearance of the surface state and renormalization of the 5$d$ band velocity. Fits of the 5$d$ band velocity at energies beneath the 4f bands (from $E=$ -150 to $E=$-50 meV) are shown in Fig. \ref{fig:fig4}(c). The bare 5$d$ band velocity obtained from overlaying theory with data on a large ($\sim$2 eV) energy scale is roughly 7.3 eV-$\AA$ \cite{SM}, and is reduced to $v_d\sim$2 eV-$\AA$ at low temperature. A factor of $\sim$2 recovery of band velocity is observed upon heating to T$\sim$150K for samples with lower doping (hollow circles), compared to an average factor of just 1.1$\pm$0.1 for both of the highly doped Ce 0.5 and Eu 0.3 samples. The trend for lightly doped samples can be understood by noting that as the 4$f$ bands gain coherence at low temperature and a topological gap opens, the level repulsion effect of 4$f$/5$d$ hybridization directly reduces the energy of 5$d$ states that draw close to the 4f band beneath the Fermi level. The effect of this on the fitted 5$d$ band velocity within the modeled spectral function [from Fig. \ref{fig:fig1}(f-h)] is plotted with a dashed curve, and accurately reproduces the convergence towards $v_d\sim$2 eV-$\AA$ in the T$\leq$100K regime. The divergence of the prediction from the fit at $T>100K$ corresponds to a region in which the 4$f$ state width is much more difficult to characterize, due to the 4$f$ states being incompletely visible beneath the Fermi level and having poor contrast on top of the incoherent background intensity.

The loss of surface states in the undoped sample occurs near T$\gtrsim$120K \cite{Jiang_2013,denlinger2014temperature,miao2020robust}, and has been associated with the increase of the f-electron imaginary self energy beyond $Im(\Sigma)\gtrsim$40 meV. The associated T$\sim$120K crossover behavior and surface state emergence are theoretically reproduced in the Fig. \ref{fig:fig1}(g-h) model, which is parameterized with self energies from Ref. \cite{denlinger2014temperature}. This $Im(\Sigma)$ value is estimated as full width at half maximum of the fitted ARPES 4$f$ band feature, which shrinks to $\sim$15 meV at low temperature. The low temperature f-electron energy width of Ce 0.3 and Eu 0.2 samples has been measured under similar conditions \cite{miao2020robust}, and has values of 36$\pm$2 and 27$\pm$3 meV, respectively. Extrapolating linearly with doping, we naively expect Ce 0.5 to be well beyond the critical value ($Im(\Sigma)\sim$50 meV), and Eu 0.3 to be just inside it ($Im(\Sigma)\sim$33 meV), consistent with the apparent decline of topological surface state visibility in Ce 0.5.

The amplitude of this energy and temperature scale for losing topological surface states is perplexingly large if we associate surface topology with the insulating gap, which is filled in due to thermalization at a much lower temperature $T_I\sim$40K \cite{HoffmanNatPhys2020} that marks the end of the exponential Kondo resistivity regime \cite{Menth_1969}. The insulating gap is consistently assigned relatively small values of $\Delta_I \lesssim 20$ meV \cite{AllenReview} ($\Delta_I \sim$12 meV in Ref. \cite{HoffmanNatPhys2020}). These factors make it important to distinguish between the hybridization gap that introduces topology and the dispersion extrema that define conductivity.

The highest energy bulk valence band features identified by ARPES are local dispersion maxima projecting to the 2D $\overline{H}$ point, which are expected to lie within a few millielectron volts of the lower edge of the insulating gap \cite{Denlinger_2013,denlinger2014temperature,HeldSmB6DMFT}. These features derive directly from the kinetic Hamiltonian of the 4$f$ $\Gamma_7$ orbital [see Fig. \ref{fig:fig5}(c)], and place a constraint on the $\Gamma_7$ orbital energy. However, recent spin texture and resonant spectroscopy analyses have revealed that the vacant 4$f$ states above the Fermi level \emph{do not} arise from $\Gamma_7$, but rather from the 4$f$ $\Gamma_8$ orbitals \cite{VojtaG8PRL, SigristG8PRL, VojtaG8PRB,SundermannG8,HeldSmB6DMFT}, and it has consequently been proposed that the topological physics should be interpreted with respect to a Hilbert space basis that neglects $\Gamma_7$. These two experimentally constraints are mutually satisfied when the $\Gamma_7$ and $\Gamma_8$ orbital energies are approximately equal, leading to a scenario in which all local maxima of the valence band emerge from the $\Gamma_7$ orbital [see Fig. \ref{fig:fig5}(a) symmetry decomposition], but the 4$f$/5$d$ hybridization that gives rise to the surface state is defined by the $\Gamma_8$ bands [see Fig. \ref{fig:fig5}(b)]. A similar band structure is achieved by recent DFT+DMFT modeling in Ref. \cite{HeldSmB6DMFT}.

In spite of the vital role played by the $\Gamma_7$ electrons in defining the insulating gap, they are effectively a spectator to the surface topology, and can be removed without qualitatively modifying the surface state dispersion [see Fig. Fig. \ref{fig:fig5}(d-e)]. The resulting picture includes a topological bulk gap of $\sim$30 meV that is commensurate to the robustness of surface states against doping and temperature, and a much smaller insulating gap. A more detailed overview of the spectator nature of $\Gamma_7$ with respect to modeling parameters, surface spin texture, symmetry mixing, and bulk Berry curvature is found in the SM \cite{SM}. A remarkable feature of this electronic structure is that the insulating gap is not primarily a hybridization gap, contrary to the conventional expectation for Kondo insulators, but can be tuned over a broad range by changing the energy difference between the $\Gamma_7$ and $\Gamma_8$ orbitals without modifying the strength of 4$f$/5$d$ hybridization. 

In spite of heat and doping playing similar roles in broadening the 4$f$ spectral features, the relative lack of doping dependence in the coherence onset temperature [Fig. \ref{fig:fig4}(b)] suggests that impurity scattering does not have a simple additive relationship with thermal decoherence. First principles-based DMFT simulations and ARPES investigations on undoped SmB$_6$ have attributed f-electron imaginary self energy as growing linearly with temperature \cite{denlinger2014temperature,SM}. Combining this with a scattering rate proportional to impurity density would yield a downward-opening `V'-shaped phase boundary [black dashed curve in Fig. \ref{fig:fig4}(b)] that is relatively inconsistent with the experimental results.

Moreover, contrary to expectations for a scenario with large 4$f$ imaginary self energy, the highly doped Eu 0.3 and Ce 0.5 samples have low temperature 5$d$ band velocities that are similar to the heavily reduced values seen in the T$\lesssim$100K Kondo coherent regime for lightly doped samples [$v_{5d}<$3 eV-$\AA$, see Fig. \ref{fig:fig4}(c)]. A likely explanation is that rather than generating incoherent scattering, impurity doping results in strong coherent local screening as suggested in recent theoretical work \cite{Baruselli_2014_PRB2,Skinner_2019}, and should therefore be thought of as contributing to real self energy rather than imaginary \cite{Anderson_SE_localization}. As such, the self energy components would not be linearly additive. An example of how this may play out for the phase boundary is plotted as a red curve in Fig. \ref{fig:fig4}(b). For this curve, the 4$f$ energy width contribution from scattering is treated as a continuum of real-valued Gaussian-distributed energies of local domains, and imaginary self energy is temperature-linear. The phase boundary is associated with 4$f$ peak width at half maximum reaching a critical threshold that corresponds to the T=120K thermal self energy. Comparing with the imaginary impurity scattering self energy scenario (black curve), the real self energy scenario has a flat-topped phase dome, consistent with the experimental data.

Characterization by ARPES in Ref. \cite{miao2020robust} found the coherence length of Eu 0.2 surface state electrons and the size scale of spatial fluctuations seen in STM to both be just $\lesssim$4 crystallographic unit cells, supporting a highly localized picture. The resistance of the topological surface state spectral features to combinations of strong disorder (as in Eu 0.3) and high temperature (T$\sim$100K) may not be surprising if the surface state spectral feature can be thought of as emerging from semi-localized single-nanometer scale domains.

%(fit Jonathan's result of the linear f band width vs temperature relation)

%Paragraph 7:
%-Broad discussion. discuss hybridization [Fig. 3(b)] and the big picture as far as coherence, scattering length scale, semi-local impurity states that theory suggests enable continued insulating character for mixed valent systems of this sort (references in our PRL***)

%Paragraph 8:
%-final summary of what was done, and the results

%suggesting that SmB$_6$ may be thought of as a parent compound from which strongly correlated topological physics can be realized with a wide range of chemical variability

In summary, we have used ARPES to map the emergence of electronic coherence phenomena as a function of doping and temperature in Sm$_{1-x}M_x$B$_6$ ($M$=Eu, Ce). Evidence of surface states is observed extending to 50\% Ce and 30\% Eu doping, raising the question of how a topological surface band structure at such high impurity density should be conceptualized, and positioning SmB$_6$ as a parent system for broad chemical exploration of strongly correlated topological physics. A crossover is found to occur beyond 30$\%$ Ce and 20$\%$ Eu doping, inducing the partial loss of surface state visibility and a significant change in the temperature dependence of 5$d$/4$f$ hybridization. Analysis of the low energy band symmetries reveals that the topological robustness can be understood by attributing the insulating and topological gaps to band pairs originating from different sets of orbitals. This has the further implication that the insulating gap of SmB$_6$ is not a hybridization gap as would be expected for most Kondo systems, and can be tuned by chemical changes that manipulate the relative energy of $\Gamma_7$ and $\Gamma_8$ orbitals. Our results shed light on the broader question of how topological insulator surface states are lost upon the incremental loss of coherence, and how the unique topological Kondo electron system realized by SmB$_6$ can be chemically tuned to explore phenomena such as the much-sought antiferromagnetic TI state that is thought to occur within Eu doped Sm$_{1-x}$Eu$_x$B$_6$ (x$\gtrsim$0.1 \cite{Yeo_2012, miao2020robust}; T$_N$ is maximized at x$\gtrsim$0.3).

%Examination of the low energy Hamiltonian reveals -- can be understood because the effective topological gap is much larger than the insulating gap; the overall scenario is interesting, because it implies free tunability of the insulating gap, to the extent that chemical changes couple asymmetrically to the $\Gamma_7$ and $\Gamma_8$ orbital energies.  ***broad doping window sets the stage for considering SmB6 not just as a strongly correlated TI, but also as the parent system for 

%The observed doping and temperature dependence are further discussed to contextualize the highly unusual resistance to impurities exhibited by the SmB$_6$ electronic structure.

% The phenomenology reveals that disorder and temperature play distinct roles with respect to the loss of topological electronic structure, and cannot be thought of as additive quantities. Our results shed light on the broader question of how topological insulator surface states are lost upon the incremental loss of coherence, and for attempts to realize the much-sought antiferromagnetic TI state that is thought to occur within Eu doped Sm$_{1-x}$Eu$_x$B$_6$ (x$\gtrsim$0.1 \cite{Yeo_2012, miao2020robust}; T$_N$ is maximized at x$\gtrsim$0.3).

This research used resources of the Advanced Light Source, a U.S. DOE Office of Science User Facility under Contract No. DE-AC02-05CH11231. Operation of the ESM beam line at the National Synchrotron Light Source is supported by DOE Office of Science User Facility Program operated for the DOE Office of Science by Brookhaven National Laboratory under Contract No. DEAC02-98CH10886. ARPES measurements at N. Y. U. were supported by the National Science Foundation under Grant No. DMR-2105081. M. S. S., B. Y.  K., J. W. L., and B. K. C. were supported by National Research Foundation of Korea (NRF), funded by the Ministry of Science, ICT and Future Planning (No. NRF2017R1A2B2008538). L. M. is supported by the National Natural Science Foundation of China (Grants No. U2032156 and No. 12004071) and Natural Science Foundation of Jiangsu Province, China (Grant No. BK20200348).

\bibliography{references}

\end{document}